\documentclass[twocolumn,a4paper,floatfix,showpacs,%
preprintnumbers,superscriptaddress,amsfonts,amssymb,amsmath]{revtex4}

\usepackage[T1]{fontenc}
\usepackage{mathptmx}
\usepackage[dvips]{graphicx}
\usepackage{url}
\usepackage{amsfonts,amssymb,amsmath}
\usepackage{subfigure}
\usepackage{url}

\makeatletter
\def\url@leostyle{%
  \@ifundefined{selectfont}{\def\UrlFont{\sf}}{\def\UrlFont{\small\ttfamily}}}
  \makeatother
\newcommand*{\ie}{\emph{i.e.}}

\renewcommand*{\vec}[1]{\ensuremath{\mathbf{#1}}}

\newcommand*{\E}{\vec{E}}

\begin{document}
\preprint{Version 1.02. Prepared for CEAS 2009 
-- European Air \& Space Conference,
$26$--$29$ October, 2009, Manchester, U.K. }

\title{%
FIRST Explorer -- An innovative low-cost passive formation-flying system
}

\author{Jan E.~S.~Bergman}
\affiliation{%
Aurora Scientific Consulting Ltd, 
2a Tudor Road, 
Hampton, 
Middlesex, 
TW12 2NQ, 
UK
}%
\altaffiliation[Also at the ]
{Swedish Institute of Space Physics,
P.\,O.\,Box 537,
SE-751\,21 Uppsala,
Sweden}%

\author{Richard J.~Blott}
\affiliation{%
Space Enterprise Partnerships Ltd., 
Bennetts Eastergate Lane, 
Eastergate, Chichester, 
W. Sussex, PO203SJ, 
UK
}%

\author{Alistair B.~Forbes}
\affiliation{%
National Physical Laboratory, 
Hampton Road, 
Teddington, 
Middlesex, TW11 0LW, 
UK
}%

\author{David A.~Humphreys}
\affiliation{%
National Physical Laboratory, 
Hampton Road, 
Teddington, 
Middlesex, TW11 0LW, 
UK
}%

\author{David W.~Robinson}
\email{david.robinson@psi-tran.co.uk}
\affiliation{%
Psi-tran Ltd., 
14 Kenton Avenue, 
Sunbury-on-Thames, 
TW16 5AR, 
UK
}%

\author{Constantinos Stavrinidis}
\affiliation{%
ESA/ESTEC, 
Keplerlaan 1, 
Postbus 299, 
2200 AG Noordwijk, 
The Netherlands
}%

\begin{abstract}
Formation-flying studies to date have required continuous and minute
corrections of the orbital elements and attitudes of the spacecraft.
This increases the complexity, and associated risk, of controlling the
formation, which often makes formation-flying studies infeasible for
technological and economic reasons. Passive formation-flying is a
novel space-flight concept, which offers a
remedy to those problems. Spacecraft in a passive formation are allowed
to drift and rotate slowly, but by using advanced metrology and
statistical modelling methods, their relative positions, velocities,
and orientations are determined with very high accuracy. The metrology
data is used directly by the payloads to compensate for spacecraft
motions in software. The normally very stringent spacecraft control
requirements are thereby relaxed, which significantly reduces mission
complexity and cost. Space-borne low-frequency radio astronomy has
been identified as a key science application for a conceptual
pathfinder mission using this novel approach. The mission, called FIRST
(\textbf{F}ormation{}-flying sub{}-\textbf{I}onospheric \textbf{R}adio
astronomy \textbf{S}cience and \textbf{T}echnology) Explorer, is
currently under study by the European Space Agency (ESA). Its objective
is to demonstrate passive formation-flying and at the same time
perform unique world class science with a very high serendipity factor,
by opening a new frequency window to astronomy. 
\end{abstract}

\maketitle
\section{Introduction}
In passive formation-flying, the spacecraft are allowed to drift
slowly so that no expensive and complex position control systems are
required to maintain the spacecraft in predefined positions. Instead,
in the passive formation-flying paradigm, the evolving orbits and the
orientations of the spacecraft are constantly monitored. By using
advanced metrology and statistical modelling methods, the relative
positions of the spacecraft can be determined with high accuracy, even
if simple ranging sensors are used. This knowledge enables the
continuous phase reconstruction of a high-performance radio
telescope aperture to be performed, while the individual constellation
satellites rotate and drift over time. 

The FIRST Explorer constellation consists of six daughter spacecraft
with radio astronomy antennas, and a mother spacecraft for science and
metrology data processing and communications. The location of the
constellation at the second Lagrange point (L2) allows for a stable,
low-drift orbit that is sufficiently far away from Earth to avoid
severe radio frequency interference (RFI), while at the same being
close enough to maintain operations using standard telemetry systems. A
novel use of credit card sized mini solar sails on the spacecraft
enables the constellation to stay within an overall mission envelope 
\cite{JENAM2009}.

\begin{figure}[b]
\includegraphics[width=\columnwidth]{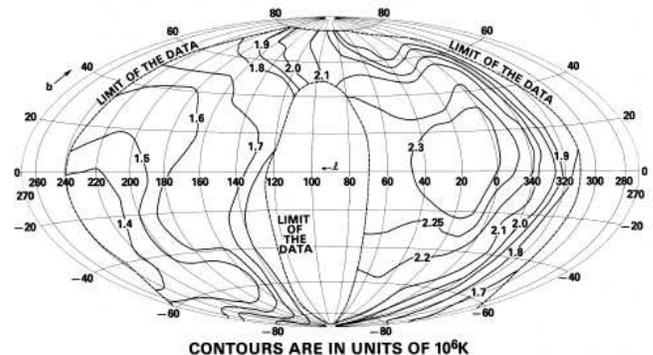}
\caption{
RAE-2 all-sky map of the galaxy at 4.70 MHz by J.~C.~Novaco
\& L.~W.~Brown (1978) \cite{Novaco78}. 
Repreduced by permission 
of J.~C.~Novaco and the American Astronomical Society (AAS).}
\label{fig:rae2}
\end{figure}

The main science objective of this pathfinder mission is to provide an
all-sky survey at very low radio frequencies, which cannot be
observed from the ground; because the ionosphere effectively acts like
a shield for frequencies below $\sim10$ MHz. Such observations have not
been performed since the Radio Astronomy Explorer (RAE 1 and 2)
missions in the 1970's, as illustrated by the all-sky
map \cite{Novaco78}
in Fig.~\ref{fig:rae2},
they were very limited in terms of
sensitivity and angular resolution. For all practical purposes, the
low-frequency Universe is therefore uncharted. A mission that could
fill this gap, and produce astronomical data of high quality, has
therefore been on the wish list of many astronomers for more than three
decades. The major science goal is therefore an all-sky survey below
10 MHz with better than 1 degree resolution. 

In addition, the FIRST Explorer can be used to image the
low-frequency Sun and the evolution and propagation of coronal mass
ejections (CME), which is important for our understanding of space
weather and its sometimes severe consequences for vital infrastructure
on Earth. It can also perform long-term studies of all the
radio-planets by providing dynamic spectra of the intense low
frequency radio bursts that originate in the planets' auroral zones.
Due to the ionospheric cut-off, those planetary radio emissions are,
with the exception of the Jovian decametric radiation, not accessible
from Earth. Similar radio emissions are most likely produced also by
extra-solar planets with magnetic fields and atmospheres. With an
intensity that can exceed their host star and a frequency content
directly proportional to the planet's magnetic field
strength, those burst radio emissions hold a great promise as beacons
for future extra-solar planet searches.

Larger space-observatories based on the enabling passive
formation-flying technology, but consisting of several small
satellite constellations, are being considered as the next step. These
observatories will have enhanced sensitivity and resolving power to
address many fundamental science objectives in radio astronomy, such as
detection and observation of extra-solar planets as well as
comprehensive studies of the dark ages and the epoch of
re-ionisation, by observations of high red-shift $21$ cm line
emissions from neutral Hydrogen, which is the only radiation believed
to have survived from the period before the Universe became transparent 
\cite{Loeb2004}.
With suitable long integration times, FIRST Explorer can in itself
gather high red-shift $21$ cm line spectra that may be able to prove
the dark ages hypothesis. A no-result is also very useful since it
will put limits on the maximum strength of these emissions at the
lowest frequencies, which are not accessible from Earth.

The FIRST Explorer mission concept study was commissioned by the
European Space Agency (ESA) \cite{FIRST} and based on an idea from one of
the authors, Dr Constantinos Stavrinidis, Head of the Department of
Mechanical Engineering at ESA/ESTEC, to reduce formation flying mission
costs. The study objective put forward by Dr Stavrinidis was to:

\emph{
``Assess the potential of an innovative and valuable science mission to
achieve with low cost, a `passive' formation flying instrument, by
allowing the formation to drift, and overcoming the need \ for
demanding and expensive technologies for position control systems to
constantly servo the various spacecraft into tightly defined
positions''.}

The cost of current and previous formation flying mission designs are
driven up by the need to \ employ sophisticated, often optical and
thereto expensive, range sensors to monitor the position and
orientation of separate spacecraft to high precision; and the need to
use complex active multi-axis micro thrusters on each spacecraft, to
constantly servo the constellation into a defined position and
orientation.

This is a challenge for a formation involving only two or three
spacecraft, but as the number increases the complexities of the servo
control algorithm grows exponentially and then the growing
time-delays between command and control make the challenge even more
difficult.

The passive formation flying concept optimises the ``knowledge'' of the
relative positions of the spacecraft -- rather than optimising control
of the position of the spacecraft. More sophisticated computational
modelling is used as a trade-off against less sophisticated
conventional ACOS (Attitude and Orbit Control System) instrumentation
using active multi-axis thrusters.

This paper examines one possible mission concept using example design
criteria, and based on existing low-cost technology that has been
demonstrated in space or on the ground. It is not a full mission design
study and many of the system parameters require further detailed design
trade-off studies, but the paper illustrates what could be achieved
as an alternative to conventional science and technology demonstrator
missions.

By adopting a different paradigm, where the instrument adapts to a
changing environment rather than the other way round and by postulating
a mission scenario in a low-drift environment, we have designed a
fresh (holistic) approach to the problem. The science goals have been
iterated with the mission architecture approach, to develop a mission
concept that both tests the low cost passive formation flying idea, and
yet still targets valuable new astronomical science results that cannot
be achieved from the ground or with conventional spacecraft designs.
The resulting mission design is known as the FIRST Explorer, as it
explores both the novel mission technology and control concepts and
points the way to larger formations that can address some of the grand
science challenges open to future space-borne low frequency radio
astronomy missions.

\section{Passive formation-flying concept and top level mission
requirements}
Formation-flying studies to date have required continuous minute
corrections to the orbital positions of the free flying satellites in 6
degrees-of-freedom (DOF). This increases the complexity and
associated risk of controlling the formation. An alternative approach
is to let the formation drift slowly and only make periodic
corrections, or re-formations of the constellation, or adaptation of
the mission, if necessary.  This technique can benefit from a dynamic
and predictive metrology modelling approach to measure the relative
positions in 6 DOF to well defined range uncertainties that are within
the tolerances of the science model, and then add corrections to the
science data to counter for the drift effects. 

It so happens that a low-frequency distributed aperture radio
telescope, with the spacecraft randomly distributed in a roughly
spherical constellation, and with sensors based on sets of 3 orthogonal
dipole antenna, can reform the image dynamically provided that the 6
DOF range and orientation information is provided to an uncertainty of
between $1/100$ and $1/360$ of the
observation wavelength. With a nominal $5$ MHz observing frequency
(wavelength of $60$ m), this uncertainty (circa $\pm 150$ mm) would be
difficult to achieve from a ranging system based on simple low-cost
omni-directional radio transponders, if only the raw performance of
the individual transponders were used.

However, as we will show, the uncertainty requirement can indeed be met
comfortably, even for simple transponders, by fusing the raw data from
three or more transponders placed on each spacecraft. That requires a
multi-lateration ranging model that computes the exact range
uncertainties and combines them in a co-operative way, which
significantly reduces the net uncertainty of the combined data. This
technology has been well developed for high-precision large-scale
three-dimensional (3D) metrology in terrestrial applications. To the
best of our knowledge, we are the first to have developed and adapted
such a precision measurement model specifically for passive
formation-flying.

The low-drift requirement was important in our choice of orbit at the
second Lagrange point (L2), because the disturbances there are
relatively modest. Even so, a major issue for spacecraft that are using
no, or minimal active control, with conventional propulsion systems, is
that they would eventually drift out of an L2 orbit, and solar pressure
would tend to separate the spacecraft over time outside a reasonable 1
to 2 year mission operation envelope.

The solar radiation pressure, which at first glance seem to have a
negative influence on the mission life-time, can however be taken
advantage of. We have proposed a spacecraft design that passively
self-stabilizes and orientates itself towards the Sun by means of
solar radiation pressure alone; and which, in addition, uses
credit-card sized mini solar sails, with active deployment, to
maintain the formation within the mission envelope for extended periods
of time. With this design concept, use of simple mission resetting
thrusters on each spacecraft might be considered only as back-up the
first time the concept is tested at L2.

Other key issues for the radio astronomy mission is the high demands on
accurate timing and synchronization, the data communication systems
($10-14$ Mbps inter-spacecraft communication link), and the on board
processing needs for the science mission; to compensate for spacecraft
drift motions and correlate the science data, which computationally is
a very heavy load.

The overall novelty of the proposed passive formation-flying mission
concept is embodied in the combination of existing technology. For the
first time brought together in a novel mission design. The key elements
that are novel in their combination are:

\begin{itemize}
\item 
The use of low-cost mobile phone type transponders, for
inter-spacecraft ranging and communications of science and range
data, together with a combined ultra{}-low mass antenna for radio
astronomy and inter-spacecraft ranging.
\item
An advanced multi-lateration position and orientation measurement
data{}-fusion model, which improves the effective accuracy of low
precision radio range sensors by up to 30 times.
\item
A novel spacecraft design that self-orientates using the solar
radiation pressure at L2 and uses mini solar sails to keep the
constellation within an overall radio astronomy mission envelope for an
extended mission life.
\item 
The use of tri-axial antennas on each daughter spacecraft
enables a full $4\pi$ steradians field of view.
The resolution and sensitivity of the
all-sky survey will increase with mission lifetime and benefit from
the slow drifts and rotations of the constellation elements.
\end{itemize}

The top-level design challenges we have encountered are discussed in
more detail in the sections below, in each case studying existing
technology or proven technology concepts. The major issues considered
are summarized as:

\begin{itemize}
\item 
Achieving the required distributed aperture sensitivity and resolution
to advance existing science, with an affordable constellation.
\item 
Significantly improving the knowledge of the inter-spacecraft
separation and orientation over what is achieved by using simple
ranging techniques only.
\item 
A radio ranging antenna/transponder and data management system
compatible with the science sensor arrangements and demands.
\item 
Orbital control of the spacecraft separation and orientation that avoids
expensive 6 DOF active control, is not vulnerable to drift forces from
gravity gradient and solar radiation pressure, and keeps the spacecraft
stable and sun pointing.
\item 
Efficient data linking and computational data fusion for both the
science and metrology data.
\end{itemize}

\section{Mission Functional and Physical Architectures.}
In order to have sufficient sensitivity and resolution, the FIRST
Explorer science mission requires a distributed sparse antenna array.
To comply with the main science objective, which technically is to
simultaneously collect radio astronomy signals from the whole celestial
sphere, vector antennas capable of resolving all three components of
the electric field vector are required. The limitation of communication
bandwidth between Earth and L2, further requires the communication of
the high bandwidth radio astronomy signals between spacecraft; so that
they may be combined and processed to reconstruct the full distributed
telescope aperture. This in turn requires knowledge of the relative
separation and orientation of the spacecraft. After the appropriate
level of on board processing, the reduced astronomy data must
subsequently be transmitted back to Earth, for further
post{}-processing, distribution and analysis.

In addition to being capable of capturing the science data to meet the
science objectives, the FIRST Explorer technology mission has to
perform within several constraints. The spacecraft must be within range
of the sensing and communicating metrology (less than $25$ km at the
limit), with sufficient inter{}-sensor visibility for determining
relative separation and orientation to the required accuracy. Solar
panels must br aligned to the Sun axis within $\pm 15$ degrees. A
sufficient mission duration is required to meet the science and
metrology objectives (one year nominal, greater than two years
preferred). Mass, power, thermal and propulsion budgets must
nevertheless be consistent with a low cost mission.

Although the constellation needs to be in a stable orbit it does not
need to be pointed in a particular direction to make science
observations. A random volume distribution is preferred, taking account
of collision avoidance and metrology and communication ranging
limitations. A planar distribution must be avoided, because in free
space a plane antenna array will have no means to tell what is ``up''
or what is ``down''. A spherical rather than disc shaped distribution
is also more helpful to the metrology. Both the science and the ranging
metrology model objectives can benefit from changes to the relative
positions of the spacecraft within the operating constraints.

In principle, spacecraft operating at L2 will experience a balance
between the combined gravitational pull of the Sun and Earth, and the
centripetal force associated with orbiting the Sun at that distance in
one Earth year. In practice, there is a gravity gradient so that a
spacecraft slightly further away will experience a lower gravitational
attraction, and a higher centripetal force from the higher speed to
complete an orbit of larger circumference. The net effect is an outward
drift \cite{XEUS} and the formula for calculating the relative acceleration
$a_{\mathrm{gg}}$, due to the gravity gradient, of two spacecraft at
L2, separated by a radial distance, $D$, is

\begin{equation}
a_\mathrm{gg}=2D\frac{\mu_\mathrm{S}+ \mu_\mathrm{E}}
{r_\mathrm{S}^3-r_\mathrm{E}^3}
\end{equation}

where $\mu_\mathrm{S}=1.327 \times 10^8$ km$^3$s$^{-2}$ and
$\mu_\mathrm{E}=3.986 \times 10^5$ km$^3$s$^{-2}$ are
the gravity coefficients for the Sun and Earth, respectively, and
where
where $r_\mathrm{S}=1.48 \times 10^8$ km and
$r_\mathrm{E}=1.5 \times 10^6$ km are
the radial distances from the Sun and Earth, respectively.

As an example: two similar spacecraft where one is $1$ km further away
from the Sun will experience a relative drift of about $150$ km in a
year. 

Spacecraft at L2 will also experience acceleration
$a_\mathrm{SR}$, away from the Sun due to solar radiation pressure,
which may be calculated from

\begin{equation}
a_\mathrm{SR}=\frac{c_\mathrm{R} S}{c}\frac{r_\mathrm{E}^2}{r_\mathrm{S}^2}
\frac{A_\mathrm{SC}}{M_\mathrm{SC}}
\end{equation}

where 
$c_\mathrm{R} = 1.6$ is the
(assumed) radiation coefficient, 
$S = 1353$ Wm$^{-2}$ is the solar
constant, 
$c$ is the speed of light, 
$M_\mathrm{SC}$ is the spacecraft mass, and
$A_\mathrm{SC}$ its effective area.

\begin{figure*}[t!]
\includegraphics[width=\textwidth]{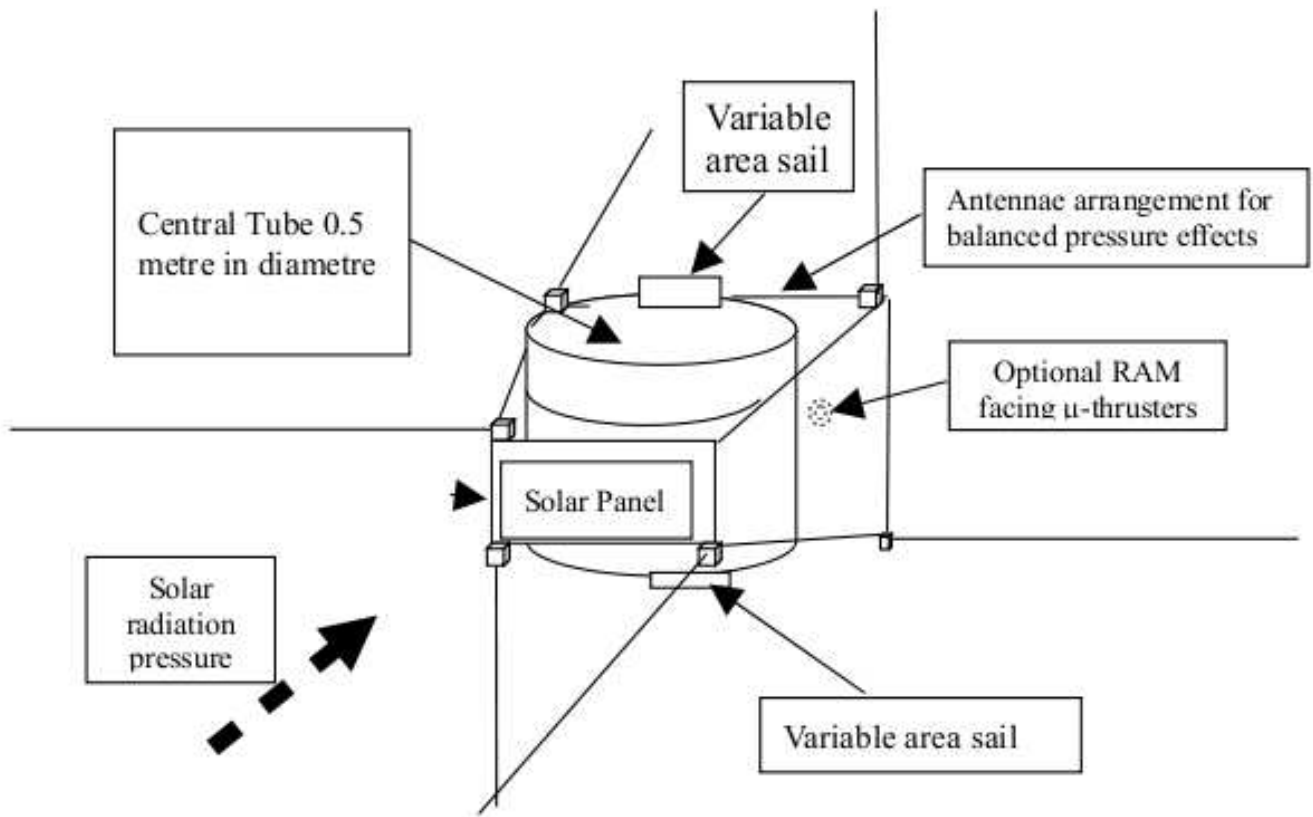}
\caption{Daughter spacecraft architecture. The central $0.5$ m diametre
tube provides a thrust tube for launch configuration and houses 
all electronics and the
variable sail area. The sun facing (front) area 
is $67\times15$ cm$^2$ and has a solar panel. The RAM face (back)
area is $88\times29$ cm$^2$. The
sloping sides, top, and bottom, with $10$ degree Chamfer and $60$ cm lengths,
provides self-correction to angular drift. The six monopole antennas 
are arranged to balance solar pressure effects. They are connected pairwise
to form three orthogonal dipoles (one vector antenna). The ranging sensors, 
located at their tips, doubles as inter-spacecraft communication
antennas.} 
\label{fig:daughter}
\end{figure*}
Spacecraft relative drift from solar radiation pressure is extremely
sensitive to differences in the effective area presented to the
pressure. For example: two $10$ kg spacecraft with effective areas $0.314$
m$^2$ with a machined tolerance of $1$ mm could experience
a difference in relative drift up to $900$ km in a year.

In theory it might be possible to select a radial
distance slightly closer to the Sun and Earth where the solar radiation
pressure was balanced by slightly stronger gravitational pull but natural
variations in solar pressure and gravitational effects make this
impractical.
Some form of drift management is therefore required to meet the 
basic mission
requirements. 

A simple, robust and highly reliable method of controlling relative
drift is by changing the effective area exposed to the solar radiation
pressure. Because relatively small differences have significant
effects over longer periods of time the changes to effective area can
be quite small. The easiest form is an extending or retracting `sail
area' as shown in Fig.~\ref{fig:daughter}. 
Depending on the design this mini solar
sail could be controlled by a space qualified `stepper motor' or even
possibly thermally actuated expansion and contraction. It would be very
important to ensure that the effect was balanced so as not to introduce
unwanted torques. A further refinement might be to effect some `limited
steering', by offsetting the mini sails and ``tacking across the solar
pressure field'', but this would need to be modelled carefully to
determine feasibility and traded-off against the risks from greater
complexity. A fall-back option is to put a micro thruster in the RAM
face to counter the rate of drift.

This approach indicates a daughter spacecraft design similar to that in
Fig.~\ref{fig:daughter} with a target mass of $10$ kg. The wedge-shaped body is built
around a central circular thrust tube, which can also be used to stack
the spacecraft for launch and transit to L2. The one metre monopoles
for LF sensing and metrology antennae deploy once the spacecraft
reaches its operational orbit. The electronics are packaged within the
thrust tube to give maximum radiation shielding. A $50$ W continuous
power budget from a $0.1$ m$^2$ array has been estimated as detailed in
Table~\ref{tab:daughterpower}.

\begin{table}
\caption{Power budget for daughter spacecraft}
\label{tab:daughterpower}
\begin{tabular*}{0.9\columnwidth}{@{\extracolsep{\fill}}|l|r|}
\hline
{\bf{Main power consumers}} & {\bf{Power [W]}} \\
\hline
\hline
Data processing 
&$5$\\
\hline
Timing (rubidium standard)&$10$\\
\hline
Metrology (inter-satellite comms.) & $12$\\
\hline
Mini solar sail area management 
&$10$\\
\hline
Propulsion (operated periodically) & $5$\\
\hline
Margin: $20$\% & $8$\\
\hline
\hline
{\bf{TOTAL}} & $\mathbf{50}$\\
\hline
\end{tabular*}
\end{table}

\begin{table}[b!]
\caption{Power budget for the mother spacecraft}
\label{tab:motherpower}
\begin{tabular*}{0.9\columnwidth}{@{\extracolsep{\fill}}|l|r|}
\hline
{\bf{Main power consumers}} & {\bf{Power [W]}} \\
\hline
\hline
{\bf{Monoevre}} (deployment and mantanance)
&$\mathbf{\le200}$\\
\hline
\hline
{\bf{Operation}} (X-band transmitter)
&$\mathbf{\le 200}$\\
\hline
Metrology (mother-daughter comms.)
&$5$\\
\hline
Timing 
& $5$\\
\hline
Data processing 
& $15$\\
\hline
Margin & $25$\\
\hline
\hline
{\bf{TOTAL}} operational requirements & $\mathbf{250}$\\
\hline
\end{tabular*}
\end{table}

\begin{figure}
\includegraphics[width=\columnwidth]{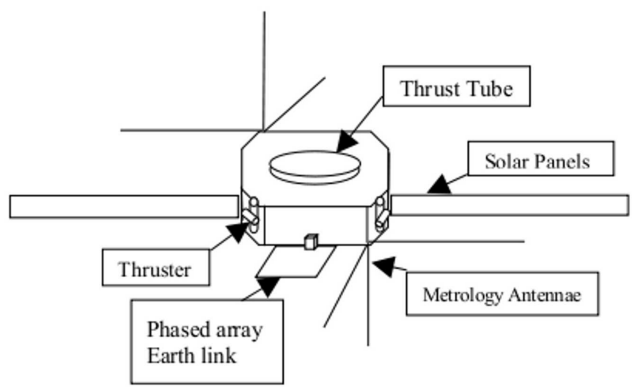}
\caption{The hexagonal three-axis stabilised mother-spacecraft is built
around a central thrust tube, $50$ cm in diameter. It has flat top panels
and two solar panel wings. Active thrusters are used for full 6 DOF spacecraft
control.
A steerable phased array is used for communication with Earth. 
The antennas used for ranging and inter-spacecraft
communication are of the same design as the vector antennae
on the daughter spacecraft. They can be used for science, providing a 
$7^\mathrm{th}$ radio astronomy antenna or be used as
a spare, if a daughter fails.}
\label{fig:mother}
\end{figure}

The Mother spacecraft will need to manoeuvre very precisely with 6 DOF
to deploy the daughter spacecraft with negligible relative linear or
angular velocity. Ideally this should vary by less than $0.1$ ${\mu}$N
of thrust or torque. It will therefore have to be highly stable during
these manoeuvres suggesting a highly effective 6 DOF AOCS and a
significantly larger mass than the daughter spacecraft. The mother
spacecraft also needs sufficient propulsion for the transit to L2 and
to remain in loose formation with the daughter spacecraft once they
have been deployed.

It is assumed that the mother spacecraft will be built around a 0.5 m
thrust tube providing the interface to the launch separation system
below and the daughter spacecraft stack above.

The most challenging task for the AOCS is the very precise control of
the separation of the daughter spacecraft. This requires precise
knowledge and control of the motion of the mother spacecraft at the
time of each separation. A star tracker, even supplemented with
sophisticated MEMS gyro arrangements, is unlikely to offer the
necessary accuracy.  The challenge therefore is to exploit the
in-built metrology range sensor network in place for the
formation-flying; in this case however the sensing will be `rate'
rather than `position' based. This may extend the deployment phase to
allow sufficient `settling time' and to build up a sufficient
statistical sample to achieve the necessary accuracy.

The mother spacecraft must be capable of independent navigation to L2
requiring Sun sensors and a simple star tracker. The sun sensors would
need to be arranged on all potential Sun facing surfaces. The star
tracker will probably be best placed on the side of the spacecraft
facing away from the sun during transit to L2. For redundancy purposes,
two star trackers are advisable. Miniature video cameras to observe
critical manoeuvres may be considered.

\begin{table}[t]
\caption{Mass budget for the FIRST Explorer spacecraft}
\label{tab:mass}
\begin{tabular*}{0.9\columnwidth}{@{\extracolsep{\fill}}|l|r|}
\hline
{\bf{Sub-system estimate}} & {\bf{Mass [kg]}} \\
\hline
\hline
Electric propulsion (wet mass)
&$50$\\
\hline
Solid chemical motor (transit)
&$30$\\
\hline
Communications equipment
& $5$\\
\hline
AOCS
&$5$\\
\hline
Power generation and distribution
& $15$\\
\hline
Timing and data processing
& $5$\\
\hline
Structure (incl. separation system)
& $45$\\
\hline
Margin & $40$\\
\hline
\hline
{\bf{TOTAL}} Mother & $200$\\
\hline
Six daughters ($10$ kg each) & $60$\\
\hline
\hline
{\bf{TOTAL}} All spacecraft & $260$\\
\hline
\end{tabular*}
\end{table}

The propulsion system must both provide the transit to L2 and precise
control the spacecraft in 6  DOF once there. Probably only electric
propulsion can offer the fine control needed but it is unlikely to also
have the power to achieve a rapid transit to L2. This probably needs a
small solid chemical motor. For the electric propulsion one can
envisage two thrusters for the $Y, Z$, pitch, roll and yaw motions and four
thrusters for the $X$ motion; each thruster is assumed to be capable of $2$
mN in the normal thrust mode and less than $0.1$ ${\mu}$N in the
differential mode. Rule of thumb calculations indicate that a $200$ kg
mother spacecraft and six $10$ kg daughters, can move the kilometre or so
needed between initial deployment positions in 4 or 5 hours before
settling time. It would therefore appear reasonable to plan the full six
spacecraft deployment sequence over 6 to 12 days.

The mother spacecraft data processing will at a minimum need to combine
the science and range metrology data streams from the daughter
spacecraft and re-transmit a pre-processed product to Earth.
Preferably, the majority of science data fusion and metrology data
processing would take place on the mother spacecraft with only the
final processed product being relayed to earth.
Even after processing there will be a reasonable volume of data to be
transmitted back to Earth and there may be instances where raw data is
required for analysis or anomaly investigation. A trade off study is
needed to establish the best balance between mother spacecraft
processing power and the volume of data for Earth transmission. At this
stage we propose an X-band link, capable of several Mbps data rate,
operating through
a flat plate phased array which can be deployed from the side of the
Mother spacecraft after separation from the launch vehicle. Some
rudimentary mechanical steering may also be needed. 

The estimated power budget for the mother-spacecraft
is based on realistic 
assumptions and detailed in Table~\ref{tab:motherpower}, which
in round terms suggests the need for $250$ W of BOL installed
power, which ould be met with two $2\times0.25$ m$^2$ solar panels. 
These considerations
suggest a mother spacecraft design concept as shown in Fig.~\ref{fig:mother}.

Given the relatively low propellant requirements with electric
propulsion, a total mass budget of $260$ kg for the 
FIRST Explorer mission
has been estimated, as 
detailed in Table`\ref{tab:mass}
It is  considered to be challenging
but not unreasonable (based on a $20$\% margin).

\section{Communications and ranging system}
The overall system aims to fulfil three objectives: transmission of the
science data, transmission of signal and control data, and ranging.
Within the approach that we have chosen, RF power, system complexity,
operating-range limits and range accuracy are issues that have
conflicting requirements. The principal technologies that we have
considered are IEEE 802.11 (WiMAX) \cite{IEEE802.11}, which is an Orthogonal
Frequency Division Multiple Access (OFDMA) $20$~MHz bandwidth system
capable of transmission rates of up to $54$ Mbps and Wideband Code
Division Multiple Access (WCDMA), which is a $5$ MHz bandwidth system
used for mobile communications with data rates of up to $384$kbps \cite{UMTS}.

In the ideal configuration, the separation of the mother and daughter
satellites will be comparable and in the region of $5-15$ km. The
optimum configuration for communication is a star network, as used in
mobile communications and wireless networks, with the mother as the
controller.  However, if the satellites drift into a non-ideal
configuration, the separation between all satellites and their near
neighbours may be close enough to permit communication but may exceed
the range limits for a star network. An alternative data path is
required in case the direct link to the mother spacecraft is lost. The
required configuration will be a peer-to-peer or ad-hoc network,
similar to mobile sensor network architectures, where one of the
daughter satellites acts as the controller, and the data is passed back
to the mother spacecraft in multiple hops. 

\begin{figure}
\includegraphics[width=\columnwidth]{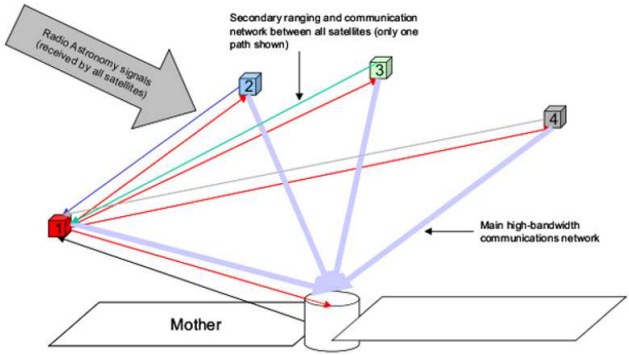}
\caption{
Communications and ranging showing the primary and secondary
networks. Two-way communication between `Daughter 1' and the other
satellites provides ranging and data communications.}
\label{fig:comms}
\end{figure}

As illustrated in Fig.~\ref{fig:comms}, the proposed communications 
architecture will comprise a main,
high-bandwidth star network centred on the mother satellite and a
secondary reconfigurable network between all of the satellites. The
secondary network also provides both communication and ranging
functions.

Within the constellation, the majority of the communication bandwidth
will be used to transmit the science data back to the mother satellite.
The science data from each of the satellites requires a capability of
circa $14$~Mbps to support an acquisition bandwidth of $100$~kHz. 

As all the spacecraft drift they will have different rotation rates and
relative orientations so it is essential that the antennas radiate with
a wide field pattern to maintain communication. The path loss can be
calculated using the Friis equation \cite{Friis}, which can be written 
\begin{equation}
P_\mathrm{r}=P_\mathrm{t}
G_\mathrm{t}(\theta_t,\varphi_t)G_\mathrm{r}(\theta_r,\varphi_r)
\left(\frac{c|\vec{\epsilon}_r\cdot\vec{\epsilon}_t^\ast|}{4\pi f D}\right)^2\,,
\label{eq:friis}
\end{equation}
where  $P_\mathrm{t}$ and  $P_\mathrm{r}$ are the transmitted
and received powers, respectively,
$G_\mathrm{t}$ and $G_\mathrm{r}$ the corresponding antenna gain 
patterns, $f$ is the frequency, $c$ is the speed of light, and  
$D$ the separation distance.
Polarisation losses
are accounted for by means of the absolute value squared of the scalar
product between the antenna polarisation (unit) vectors, 
$\vec{\epsilon}_\mathrm{t}$ and
$\vec{\epsilon}_\mathrm{r}$. These are in general complex valued 
but becomes real valued for linearly polarised antennas.

For the calculations
we have assumed an antenna gain of $1.5$~dB, typical of a linearly
polarised electric dipole antenna,
and average orientation and polarisation penalties of $3$~dB. At $10$~km
range and an operating frequency of $2$~GHz we anticipate an average path
loss of $121$~dB between any pair of antennas. In practice, as there are
several antennas to allow for the relative orientations of the
satellites it will be possible to make use of receive (and transmit)
diversity to increase the system sensitivity.

Power, range and data throughput are intimately connected. Previous work
by Mahasukhon \emph{et. al.} \cite{Puttipong} showed that the
popular IEEE 802.11 WLAN standards, require a Signal-to-Noise Ratio
(SNR) of about $22$~dB to achieve $54$~Mbps operation and $3$ dB to achieve 
6 Mbps operation. Although the maximum bit rate is $54$~Mb/s, the
saturation throughput depended on the exact implementation of the
standard. For a channel bit error rate (BER) of $10^{-5}$ the saturation
throughput is $23.2$~Mbps for IEEE 802.11a and $9.7$~Mb/s for IEEE 802.11g.
The inference is that, the Media Access Control (MAC) layer and
operating frequency will need to be tailored for this application to
achieve the required performance and range.

\begin{figure}[b!]
\includegraphics[width=\columnwidth]{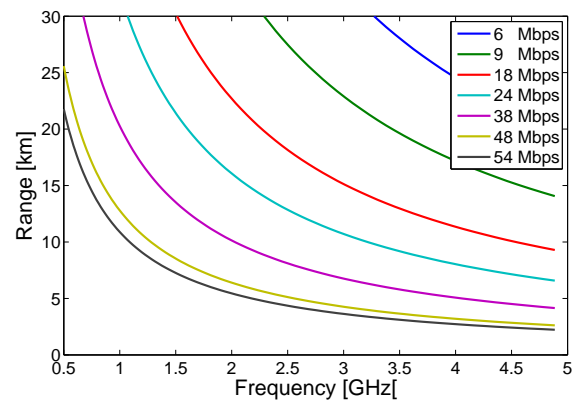}
\caption{
Data rate range and operating frequency
limits for WLAN. The SNR values ($n_\mathrm{SMR}$) employed,
from $6$ Mbps (top curve) to 
$54$ Mbps (bottom curve), were, 
$3.3, 6.4, 10, 13, 17, 21$, and $22.4$.}
\label{fig:rlim}
\end{figure}

A noise power of $-100.8$~dBm has been calculated assuming a $20$~MHz
bandwidth and a receiver temperature $T_\mathrm{rec}=300$~K. The
limiting range can be found for any frequency and SNR by rearranging
the Friis equation, Eq.~(\ref{eq:friis}), and using the noise 
threshold levels calculated from
\cite{Puttipong}. An additional factor, $G_\mathrm{d}$, which accounts for
the diversity gain, has been
included in the equations because each satellite uses several antennas
and these signals can be combined constructively to increase the
sensitivity. In this analysis the average value of
the diversity gain has been conservatively set
to $G_\mathrm{d}=2$ ($3$ dB). The range limit can then be calculated from
\begin{equation}
D_\mathrm{lim}=
\frac{c|\vec{\epsilon}_r\cdot\vec{\epsilon}_t^\ast|^2}{4\pi f}
\sqrt{
\frac{P_t\langle G_\mathrm{t}\rangle\langle G_\mathrm{r}\rangle 
\langle G_\mathrm{d}\rangle}
{k_\mathrm{B} T_\mathrm{rec} n_\mathrm{SNR}\Delta f}}\,,
\label{eq:Rlim}
\end{equation}
where  $k_\mathrm{B}$ is the Boltzmann constant,  $\Delta f$ is the bandwidth, 
$\langle G_\mathrm{t}\rangle$ and $\langle G_\mathrm{r}\rangle$
are the average values for the
antenna gains (isotropic, 0 dBi) and  $n_\mathrm{SNR}$ is the signal-to-noise
ratio (SNR) required to support a specific data rate. 
The polarisation factor 
$|\vec{\epsilon}_r\cdot\vec{\epsilon}_t^\ast|$ has been set to $1/\sqrt{2}$,
which corresponds to an average misalignment between the polarisation vectors
of $45$ degrees.
The results,
shown in Fig.~\ref{fig:rlim},  infer that a system based on IEEE 802.11 coding standards
will meet the bandwidth and range requirements.

The ranging system provides the data for the multi-lateration model to
determine the relative positions and orientations of the spacecraft.
The two main ranging options are to use an independent,
self-contained solution, or to use a cooperative approach. The
limitation of the self-contained radar approach is that the returned
power is inversely proportional to the fourth power of the separation
between the source and target, limiting the operating range. For this
reason a cooperative, communications based approach has been adopted.
Each two-way communication link provides ranging information so the
desired outcome is to maintain communication links between all or as
many of the satellites as possible. The number of two-way links 
$N_\mathrm{2way}$  can be calculated from the number of satellites  $N$ as
$N_\mathrm{2way}=N(N-1)/2$.

The system will require many separate communication links. Code Division
Multiple Access (CDMA) techniques use orthogonal codes to provide
diversity and hence allow the re-use of the carrier frequency. A code
length of 256 bits will provide 64 orthogonal code channels.

In ground-based systems, the location accuracy of mobile transmitters
within a cellular radio system may be limited by non line-of-sight
errors \cite{CDMA} to about $20$ metres and this can be considerably improved
using interferometric techniques to give accuracies of $0.6$ m
\cite{Tracking}. In a space environment, multi-path
fading signals will be much less of an issue as the separation and
cross-section of the satellites is small. There will be some
interaction between the antennas and the spacecraft that will require
calibration and correction to achieve the required accuracy.

\begin{figure}[b]
\includegraphics[width=\columnwidth]{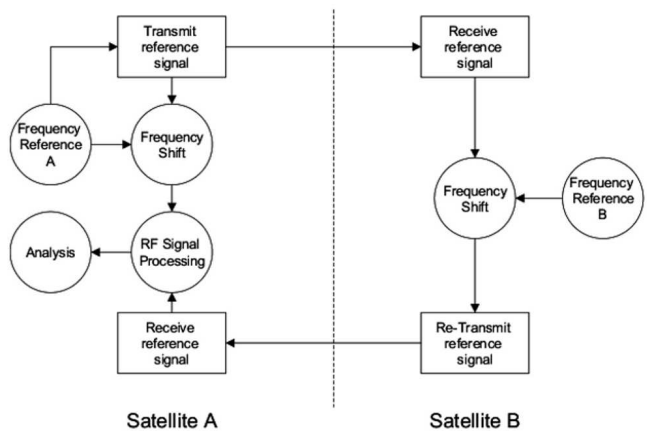}
\caption{
Cooperative ranging by re-transmitting a frequency-shifted
signal.}
\label{fig:ranging}
\end{figure}

The inference of these studies is that the timing accuracy achievable
through the MAC layer is insufficient to provide the required ranging
accuracy but will ensure that more accurate methods do not suffer from
cyclic errors.

Several options are available for cooperative ranging: time of flight
for a re-transmitted signals gives a resolution that is inversely
proportional to the signal bandwidth $\Delta R\approx c/3\Delta f$.

Edge timing of a Time-Domain Diversity communications system with a
bandwidth of $20$~MHz would correspond to an accuracy of $5$~m, which is
insufficient for the requirement. Using a short RF pulse, rather than
the communications system, could improve the accuracy by an order of
magnitude but co-operative management of the ranging function
throughout the constellation would be more difficult as only one
satellite transponder could obtain range information at a time.

The relative motion of the satellites must be taken into account in any
implementation. As the distance between the two craft separates the
phase of the received and demodulated RF signal will rotate through
$2\pi$ for every wavelength of separation. In a typical system, the
phase or frequency of the Voltage Controlled oscillator will track
these variations. If this information can be accessed within the RF
chip/chip set it will correspond to an uncertainty of circa
$\pm30$~mm for a $2$ GHz carrier, assuming a phase stability of
$\pm1.25$ radians.

Consider the simple implementation shown in Fig.~\ref{fig:ranging},
where a received signal is frequency
shifted and retransmitted. As the reference oscillators within each of
the satellites are independent, the signal processing system must be
designed to separate frequency shift, velocity components, and $2\pi$
ambiguities. Uncertainties in satellite time-transfer have been
studied in detail \cite{Instabilities}, and the expected uncertainty of the transponder
path-delay calibration is expected to be less than $1$~ns \cite{Whibberley}, giving
an overall uncertainty better than $\pm150$~mm.

In summary, a communication system based on COTS hardware with
additional components is expected to have the capability of providing
the required data handling ($14$~Mbps per satellite) at circa $5-25$ km
separation and a ranging uncertainty of less than  $\pm150$~mm,
as required by the proposed application.

\section{Fusing range measurements to determine formation configuration
geometry}

The basis of the FIRST Explorer passive formation{}-flying concept is
that of allowing movement between spacecraft but with the relative
locations and orientations of the spacecraft determined to precise
limits from range data. A mathematical (multi{}-lateration) model and
associated software have been developed to analyse various designs and
to predict the uncertainty in configuration geometry from range and
time measurements. 

The model assumes a fixed number of spacecraft undergoing approximately
constant linear and angular motion where each spacecraft has a fixed
number of range transponders or \textit{targets} at known positions. At
any given time, six parameters are required to determine the position
of a spacecraft: three to determine location and three to determine
orientation. Six constraints need to be applied to fix the frame of
reference of the overall constellation of spacecraft. For spacecraft
undergoing constant linear and angular motion, twelve parameters are
required per spacecraft, six to determine an initial position, three to
specify the direction and speed of the linear motion, two to determine
the axis of rotation and one to specify the speed of angular rotation.
Nine constraints need to be applied, six to fix a frame of reference
and three to specify the initial linear motion applying to the
constellation as a whole.

In practice, the behaviour of the range transponder will depend on the
angles of the line-of-sight. The model assumes that this behaviour
can be specified in terms of one or more direction vectors and
acceptance angles associated with the target position. If the angles
between line-of-sight joining two targets are smaller than the
acceptance angles at each of the targets, then the associated range
measurement is admissible. At any given time, the acceptance angles
associated with a range measurement between any pair of targets can be
estimated from the parameters describing the motion of the
constellation.

It is assumed that each measured range value estimates the distance
between a pair of targets at a given time, one target on one
spacecraft, one on another.  To each range measurement is associated a
time and two indices specifying the two targets. Uncertainties are also
associated with the range measurements, based on a characterisation of
the transponders and can be used to provide weights for the
measurements. For example, each range measurement can be weighted
according to the angles of incidence, with the maximal weight equal to
one if the line of sight is aligned exactly with the corresponding
direction vectors.

\begin{figure}[t]
\includegraphics[width=\columnwidth]{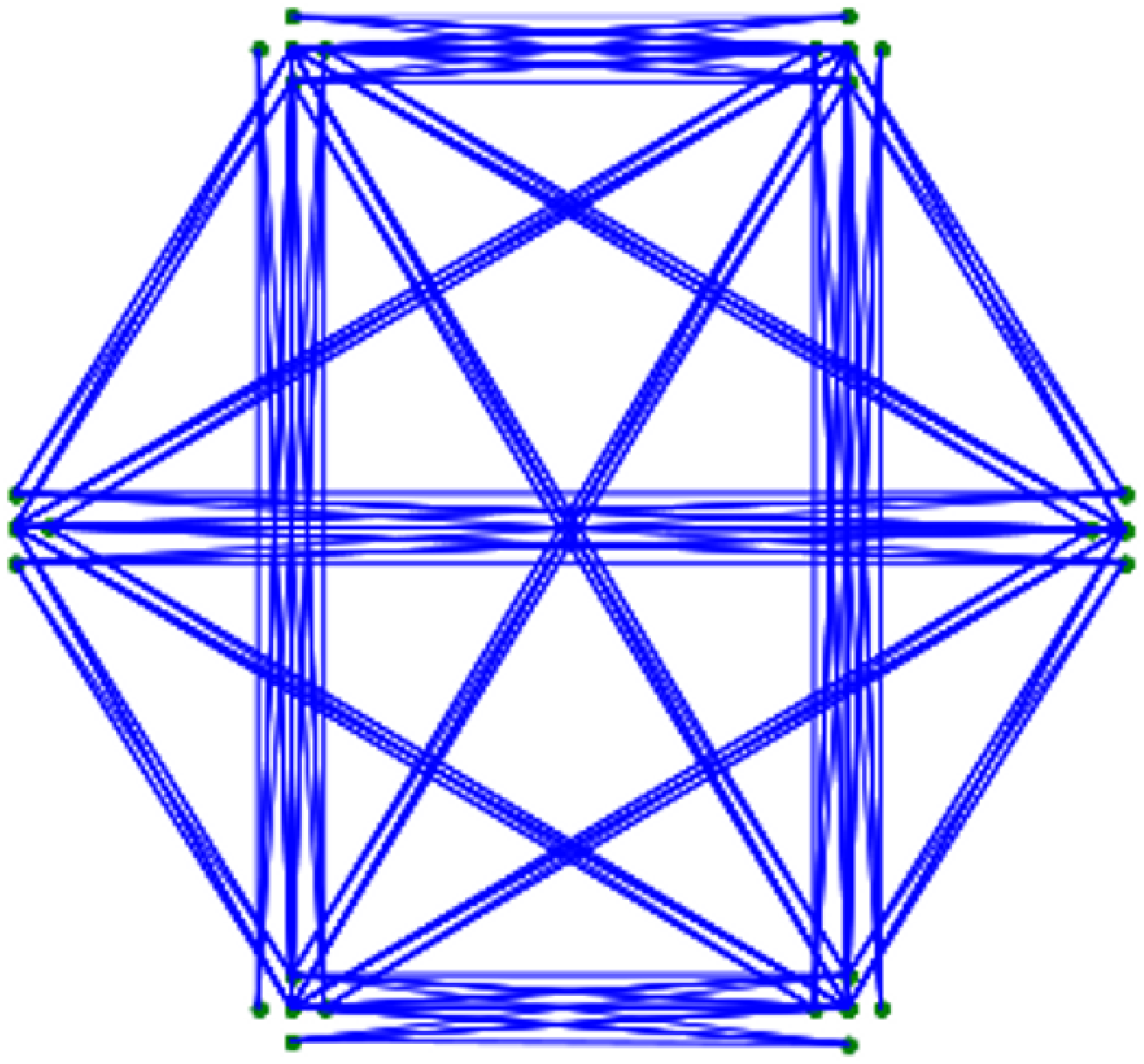}
\includegraphics[width=\columnwidth]{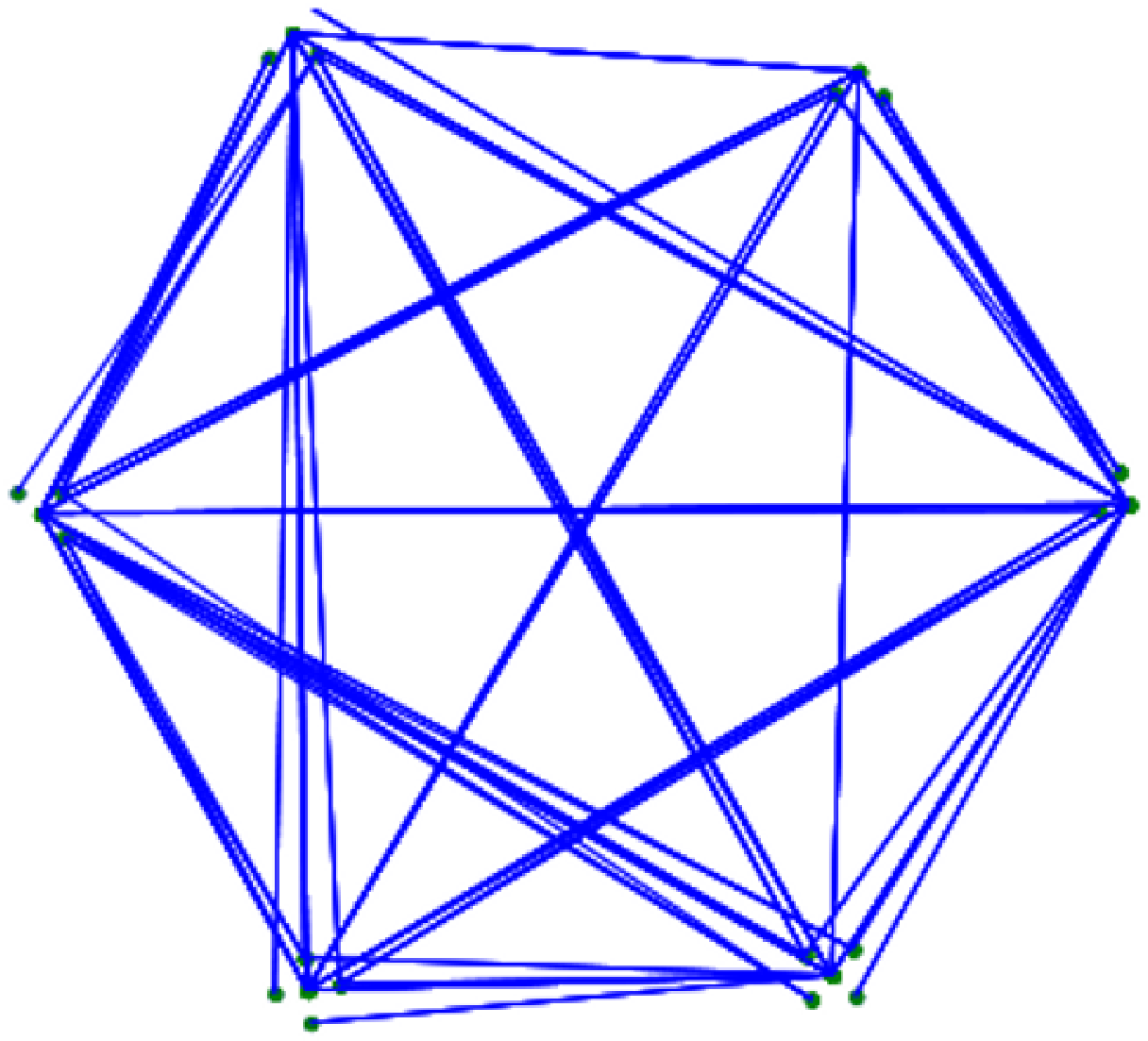}
\caption{
Admissible range measurements associated
with six targets on each of six satellites in an initial position
(top panel) and after drifting (bottom panel).}
\label{fig:range}
\end{figure}

Given motion parameters, contained in the vector $\vec{a}(t)$, 
specifying the motion of the
constellation, at any given time $t$, the distance $D_{nm}=D_{nm}(\vec{a}(t))$
between any two targets, $n$ and $m$, can be predicted. 
Given range measurements, $d_{nm}$,
estimates of the motion parameters, $\vec{a}$ can be determined by
matching the model predictions to the actual observations in the least
squares sense. The uncertainties associated with the range measurements
can be propagated through to those associated with the fitted
parameters using standard methods \cite{Forbes-1, Forbes-2, Peggs}

A set of simulation tools has been implemented in Matlab. The software
is completely flexible with respect to the number and configuration of
spacecraft, the number, location and acceptance behaviour of the
targets (range transponders on the spacecraft), and the number and
timings of the range measurements; all that is required is that each
range measurement has an associated time and indices specifying the
targets involved.

In terms of design considerations for the FIRST Explorer concept, the
number of spacecraft has been fixed at six, with up to six transponders
on each daughter spacecraft. A primary concern is having enough range
measurements to provide an accurate assessment of the constellation
geometry for almost arbitrary motion of the spacecraft. A balance has
to be made between
\begin{itemize}
\item[a)] maximising the number of admissible range
measurements for a particular, ideal constellation geometry, and
\item[b)]
making sure that if the constellation drifts from its optimal shape
there are sufficient range measurements available to meet the accuracy
requirements.
\end{itemize}

It is assumed that any constellation will suffer from drift and that
over a long period there can be no guarantee that a constellation will
match a preferred geometry. Furthermore, the slow rotation of each
spacecraft about its main axis cannot be ruled out. This means that a
transponder design that depends significantly on a particular
constellation geometry runs the risk of degraded performance as time
progresses. The simulations discussed below make the following
assumptions: the centre of gravity of each spacecraft drifts linearly a
few metres over the time scale of a few days, each spacecraft rotates
about an axis a few times per day, the rates of rotation about the axes
are different. The simulations reported below also assume that each
spacecraft maintains at least a rough alignment with the sun (up to $30$
degrees misalignment). The most significant assumptions above relate to
rotation since they are used to make sure the constellation geometry
does not become stuck in a static case in which there are very few
admissible range measurements. The conclusions below hold for different
rates of drift and rotation if the time scales are adjusted
appropriately.

The transponders have the following acceptance behaviour consistent with
a radio transponder that is not fully omni-directional but has
preferred gain patterns of sensitivity. Each transponder is associated
with a direction vector. If a line of sight makes an angle of between
$60$ and $120$ degrees with the direction vector, the range measurement is
admissible for that transponder. Thus, a range measurement is
admissible for a target if it makes angles of less than $30$ degrees with
the plane orthogonal to the direction vector; this is termed a
``doughnut'' acceptance profile.  Additionally, a range measurement is
admissible for that target if the line of sight makes angles of less
than $30$ degrees with the direction vector, giving a ``one-sided
cone'' acceptance profile. A range measurement is admissible if it
satisfies at least one of the acceptance criteria at each end.

The advantage of the design of transponder is that if the direction
vector is not aligned with the axis of rotation of the spacecraft, the
doughnut acceptance region sweeps through the whole of 3D space in one
rotation so that the transponder has a chance to see all others in the
constellation. The different rates of rotation are used to guarantee
that the rotations are not locked together making some, possibly many,
of the  range measurements permanently inadmissible.

\begin{figure*}[t!]
\subfigure
{
\includegraphics[width=\columnwidth, height=7cm]{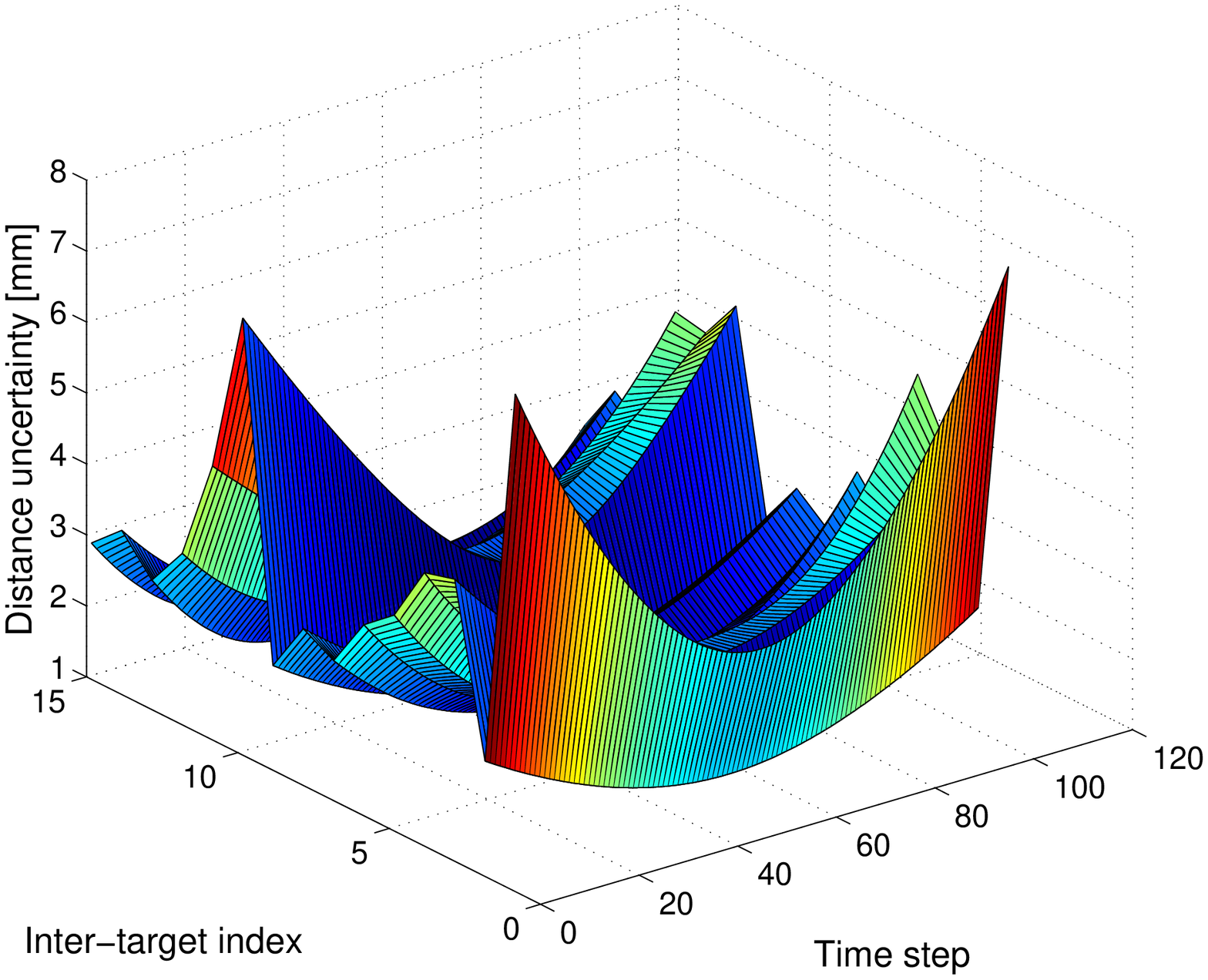}
}
\subfigure
{
\includegraphics[width=\columnwidth, height=7cm]{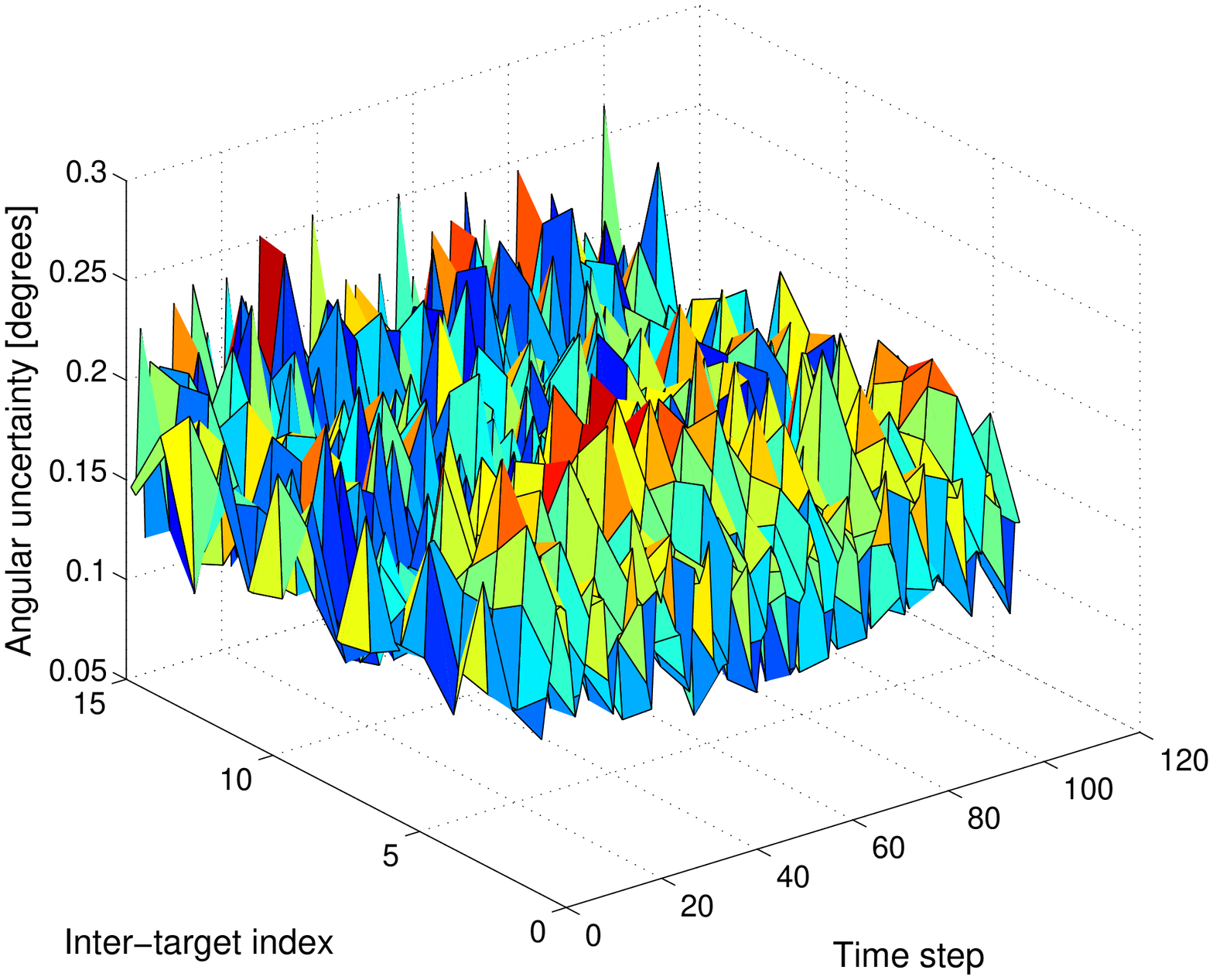}
}
\caption{
Uncertainties associated with the
separation distances (left panel) and angles (right panel)
between spacecraft centres of gravity at 101 time steps.
There are six spacecraft and 15 inter-spacecraft distances.}
\label{fig:uncertainty}
\end{figure*}

The following simulations involve six spacecraft randomly situated in
an oblate sphere of equatorial diameter $5$ km and polar diameter $3$ km.
Each suffers constant linear drift in a random direction and constant
rotation about a random axis aligned to within $30$ degrees of a fixed
axis (pointing to the sun). Each satellite has six transponders
situated at the centre of the faces of a cube of side $3$ m. The
direction vectors associated with the transmitter/receivers are aligned
with the vector joining the centre of the cube to the corresponding
faces.

The simulation calculates,
\begin{itemize}
\item[a)] the uncertainties associated with the
distances between the centres of gravity for each pair of spacecraft at
each time step,
\item[b)] the uncertainties associated with the angles between
fixed vectors for each pair of spacecraft at each time step, and 
\item[c)] the
uncertainties associated with the parameters specifying the constant
motion of the constellations. 
\end{itemize}

The measurement uncertainty for the
determination of the range between any two 
targets, with the radio range transponder discussed in Section 4
above, is taken to be $\pm150$ mm.

The simulations show that the ratio of admissible range measurements to
all possible range measurements is low, of the order of $5$\%. However,
as shown by the illustrations in Fig.~\ref{fig:range},
the admissibility profile can change as the
formation evolves with time. The simulation has shown that
For 100 time steps, there are typically over
5000 admissible range measurements. 

The evaluated uncertainties associated with the inter-spacecraft distances,
as shown in the left panel of Fig.~\ref{fig:uncertainty}, 
are of the order of better than $\pm10$ mm at 
all time steps 
and the uncertainties associated with the angles, as shown in 
the right panel of Fig.~\ref{fig:uncertainty}, 
are less than $\pm 0.5$
degrees. 

The uncertainties for the inter-spacecraft
distances have a typically U-shaped behaviour with respect to time.
The distances for middle time periods use information from the past and
future, while initial and final time periods can only use future or
past measurements, respectively. The uncertainties associated with the
angles are much more sensitive to which inter-target pairs are
associated with admissible range measurements and hence vary from time
step to time step as the admissibility pattern changes. Integrating
over a long time or taking measurements at a greater rate would reduce
the uncertainties further.

\section{The science mission requirements}

The Universe at very low frequencies is virtually unexplored, as
illustrated in Fig.~\ref{fig:rae2}  by the very sparse all-sky 
map at 4.70 MHz. It
was constructed from observations made by the Radio Astronomy Explorer~2
(RAE-2) \cite{Alexander75} and published in 1978 by Novaco \& Brown
\cite{Novaco78}.
It is still the best picture we have of the Universe at this
very low frequency. The main reason for our evident lack of knowledge
is the Earth's ionospheric plasma, which effectively
blocks out radio frequencies below the ionospheric cut-off, about $10$
MHz, and makes observations below $\sim30$ MHz very difficult. Therefore,
astronomical observations at low frequencies require a space borne
radio telescope.

The FIRST Explorer is aimed to be such a telescope. Its primary science
objective is to provide a new all-sky survey at very low radio
frequencies with improved sensitivity and an angular resolution
better than $1$ degree. A
secondary objective is to observe radio sources in the solar system,
such as the low-frequency Sun and non-thermal planetary radio
emissions. Those emissions will in any case interfere with 
the all-sky survey and therefore needs careful monitoring.
The secondary objective thereby
complements the primary one. A third objective is to try to detect 
high red
shift 21 cm Hydrogen line emissions \cite{Loeb2004}, or at least
put limits on
the maximum strength of the low-frequency part of this radiation, which should
now be in the $9$ --$200$ MHz range.
It is the only known way to probe the
``dark ages'' from recombination to reionisation,
a transition
period when the first galaxies were formed and
the Universe went from being completely opaque (the dark ages)
to becoming transparent when the first stars formed and were able to
ionise
the interstellar medium.
A detection could prove the dark ages hypothesis, which 
would be major step in observational cosmology.

Low-frequency radio astronomy tends to be very suitable for a space
mission. As one first might think, it does not require a huge dish
antenna to be deployed or constructed in orbit. Instead, and as we will
show, a relatively modest array of simple dipole antennas can be used.
The physical reason for this is the effective collecting area of the
dipole, which is proportional to the wavelength squared. To increase
the area, and hence the sensitivity, one just adds more dipoles to form
an antenna array; although, a space array requires several spacecraft
flying in formation. 

The sensitivity increases linearly with the number of spacecraft so
even though the  RAE-2 used a 229 meter long travelling-wave V-antenna
as well as a 37 meter long dipole \cite{Alexander75}, a space
array would be a considerable step forward. The largest improvement,
however, would be if the antenna signals could be
combined and used to perform radio interferometry. With suitable
separation distances that would dramatically improve the angular
resolution of the observations.

The science cases for space borne low-frequency radio astronomy are
well established \cite{Kassim_Weiler90} and have been worked upon 
since the last attempt by
RAE-2 and free-flying as well as lunar-based observatories
have been proposed. The Astronomical Low-Frequency Array (ALFA) \cite{ALFA}
and
the Solar Imaging Radio Array (SIRA) \cite{SIRA}, 
are two noteworthy NASA studies
of the free-flying concept, none of which became realised. Sadly, the
high cost, associated with the complex technologies traditionally
considered necessary for precision formation-flying, has always been
an Achilles heel for previously proposed low-frequency space arrays;
not to mention the cost of a deploying an antenna array
on the Moon. The FIRST Explorer, on the other hand, uses passive
formation flying, which for the first time can make a low-frequency
space observatory affordable. 

The slowly drifting and rotating
spacecraft put new requirements on the science payload, \ie, the
radio astronomical antennas.
The drift is actually not a problem but an advantage. It allows
sampling of more baselines and hence gives better coverage than would a
traditional formation, where the spacecraft are forced to stay put at
locked positions. The rotations can be taken care of by using a
\emph{vector antenna}, consisting of three orthogonal electric dipole
antennas, which allows sampling of the full three-dimensional
electric vector field, $\E(\vec(x)_n,t)$, at the satellites' positions
$\vec{x}_n$ 
and  time $t$. The electric field vector $\E_n(t)=\E(\vec{x}_n,t)$
picked up by the antenna on the $n^\mathrm{th}$ satellite 
is low-pass filtered, digitised, bandpass filtered and converted to
base-band by the radio receiver. It is represented as a time series of
complex vectors. 

Since we are only considering astronomical source the
intensities $I_n=\langle\E_n\cdot\E_n^\ast\rangle$ 
(brackets mean time average) are equal on every
spacecraft. 
Because $I_n$ is a scalar it is invariant under
rotations \cite{Carozzi2000}, also those performed by spacecraft. 
In terms of intensity, a
receiving vector antenna is therefore isotropic.

However, intensity measurements are not sufficient to perform
interferometry. We must also be able to measure the time differences
between the field vectors registered on different spacecraft. 
Let $\E_n=\E_0\exp(\mathrm{i}\Phi_n)$, where $\Phi_n$ is the phase.
The operation $\E_n\cdot\E_n^\ast=\E_0\cdot\E_0^\ast$  makes the phase
information vanish but
if we instead take the scalar product without complex conjugation (denoted by superscript $\ast$) we find that the operation 
$\E_n\cdot\E_n=\E_0\cdot\E_0\exp(\mathrm{i}2\Phi_n)$ 
preserves the phase, which can then be recovered, since
$\Phi_n=\mathrm{Arg}[\E_n\cdot\E_n]/2$.
The complex scalar $\E_n\cdot\E_n$
is also rotation invariant and in that sense the phase
$\Phi_n$ can be said to be absolute.  

If we
instead would use the phase from only one $\E$-field component
that would
not have been the case since the component values are dependent on
the coordinate system. Therefore, rigorous tracking of the spacecraft
attitude, and continuous compensation for their rotations in software,
would have been required. 

In addition to being rotation invariants,
using the Maxwell equations, it is not difficult to show that
\textbf{E}{\textperiodcentered}\textbf{E*} as well as
\textbf{E}{\textperiodcentered}\textbf{E} obey conservation laws and
are thus true physical observables. 

Having established that vector antennas can be used as isotropic and
phase preserving astronomy sensors it is clear that the performance of
the FIRST Explorer array to a very large extent will depend on precise
time and precision range measurements. 

\begin{figure}[t]
\includegraphics[width=\columnwidth]{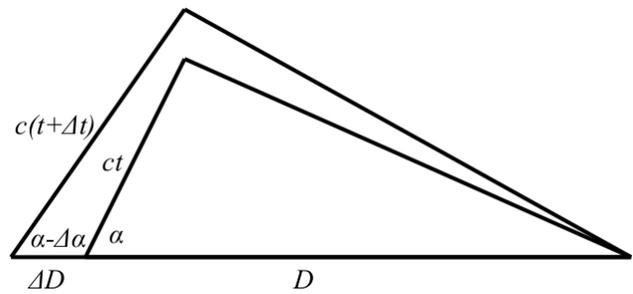}
\caption{
Geometry of a two-element interferometer of length $D$, receiving 
a radio wave from
an angle $\alpha$. The time and distance errors, $\Delta D$ and $\Delta t$,
respectively, give rise to
a pointing error $\Delta\alpha$.}
\label{fig:triangle}
\end{figure}

For simplicity, consider a two-element interferometer 
with antennas separated by a
distance $D$ and receiving a radio pulse
from an angle $\alpha$ with respect
to the baseline between the two antennas.
As illusttraded in Fig.~\ref{fig:triangle}, the radio pulse, 
which travels with the speed of light, $c$,
hits the first antenna at time
$t=0$. It will hit the second antenna at a later
time $t$. Using basic trigonometry we obtain the formula
\begin{equation}
ct=D\cos\alpha\,.
\label{eq:alpha}
\end{equation}

Hence, by measuring time $t$ and separation distance $D$ we
can calculate the direction to the source. Or vice versa, by using 
time delays
or changing the separation distance we can point the interferometer 
in different
directions. In reality, time and distance can never be determined
exactly and the errors
will result in an uncertainty in the pointing accuracy,
$\Delta\alpha$, which is the angular resolution of the
interferometer. 

Let $t\rightarrow t+\Delta t$ and $D\rightarrow D+\Delta D$, where
$\Delta t$ denote the timing error and
$\Delta D$ the error in distance, as shown in Fig.~\ref{fig:triangle}
Trigonometry
yields 
\begin{eqnarray}
c(t+\Delta t)=D\cos\alpha+c\Delta t=
(D+\Delta D)\cos(\alpha-\Delta\alpha)\,.
\label{eq:exact-error}
\end{eqnarray}

Note that 
$\Delta t>0$
gives $\Delta\alpha<0$ but $\Delta D>0$ gives $\Delta\alpha>0$, which means
that the
errors may cancel each other. To investigate if they do we assume
the errors to be independent and  Gaussian with zero mean values. 
Squaring Eq.~(\ref{eq:exact-error}) and taking the time average yields
\begin{eqnarray}
&&\langle(D\cos\alpha+c\Delta t)^2\rangle=D^2\cos^2\alpha+c^2
\langle\Delta t^2\rangle\nonumber\\
&=&\langle (D+\Delta D)^2\cos^2(\alpha-\Delta\alpha)\rangle\nonumber\\
&\approx&
D^2\cos^2\alpha
+\langle\Delta D^2\rangle\cos^2\alpha
-D^2\cos(2\alpha)\langle\Delta\alpha^2\rangle\,,
\label{eq:squared-error}
\end{eqnarray}
where in the last step we have used the Maclaurin series expansions
of $\cos\Delta\alpha$ and $\sin\Delta\alpha$. Thus, on average,
the standard deviations of the time and distance errors help
to reduce the standard deviation in pointing accuracy.
\begin{eqnarray}
\langle\Delta D^2\rangle\cos^2\alpha
-c^2\langle\Delta t^2\rangle
\approx
D^2\cos(2\alpha)\langle\Delta\alpha^2\rangle\,,
\label{eq:squared-error2}
\end{eqnarray}

A space array must be able to handle radio waves coming from all
directions. Assuming that the sources are uniformly distributed
we integrate Eq.~(\ref{eq:squared-error2}) from $\alpha=0$
to $\alpha=2\pi$, which yields 
$\langle \Delta D^2\rangle\approx 2c^2\langle \Delta t^2\rangle$.
In other words, to minimise the average standard deviation of the
pointing accuracy, the standard deviations of the time 
and distance errors, $\sigma_{\Delta t}$ and $\sigma_{\Delta D}$
respectively, should be chosen such that 
$\sigma_{\Delta D} = \sqrt{2}c\sigma_{\Delta t}$. By this choice,
the angular dependency in Eq.~(\ref{eq:squared-error2}) actually vanishes,
instead
\begin{equation}
\langle\Delta\alpha^2\rangle
\approx\frac{\langle\Delta D^2\rangle}{2 D^2}
=\frac{c^2\langle\Delta t^2\rangle}{D^2}\,.
\end{equation}

In Section V it was shown that an inter-spacecraft distance uncertainty
of $\pm 10$ mm was possible to achieve. To match that error,
as to minimise the average pointing uncertainty, would require
a time uncertainty of $\pm 24$ ps. For a $1000$ m baseline the
corresponding pointing uncertainty is only $\pm 2$ arc-seconds,
well below the required $1$ degree resolution. While
it has shown possible to achieve such small time uncertainties in ground based
antenna arrays, using a common reference frequency and adjusting
the phase of the receivers by using GPS clocks \cite{Gustav2009}, that approach is impractical
for a space observatory, which aims to be omni-directional.
Because, to make use of high time precision, one has
to delay the signals by fractions of the sampling period, which is about
$10$ ns for a state-of-the-art $100$ Msps $16$ bit ADC (analogue-to-digital 
converter).
Although it is possible to implement several fractional sample delays (FSD),
their number is finite, which results in a telescope having only a finite
number of looking directions. However, this approach may be of interest for
observation of specific radio sources, such as auroral kilometric radiation
(AKR) generation regions,
with ultra-high angular resolution.

Given the $1$ degree requirement and the $\pm 10$ mm distance uncertainty it is
clear that the required time uncertainty can be reduced significantly.
Setting $\Delta D=0$ in Eq.~(\ref{eq:exact-error}), squaring, time 
averaging and integrating over $\alpha$ yields
\begin{equation}
\langle\Delta\alpha^2\rangle
\approx\frac{2 c^2\langle\Delta t^2\rangle}{D^2}\,.
\label{eq:appox-error3}
\end{equation}
Based on the assumed sampling period we should at least have  
$\sigma_{\Delta t}=\sqrt{\langle\Delta t^2\rangle}\le10$ ns. 
Eq.~(\ref{eq:appox-error3}) then gives a separation distance 
$D\ge 240$ m for a $1$ degree angular resolution.

The above discussion is valid for observations times during which
the drift of the clocks can be neglected. For long term observations
the $\langle\Delta t\rangle=0$ assumption is no longer true. Assuming 
linear drift rate $k$ we let $\Delta t=k t+\epsilon(t)$ where $\epsilon(t)$
is the previously assumed Gaussian error. The variance of the time error
at time $t$ is then $\langle \Delta t\rangle=kt/2$. In order to perform
interferometry the accumulated time error must be less than one period
of the observation frequency, 
or the fringes will be smeared out. Hence,
$\langle \Delta t\rangle=k\Delta T/2\le 1/f_\mathrm{obs}$ is required. Here
$\Delta T$ is the observation (integration) period.
For example, observation at $5$ MHz during one year requires a drift rate 
$k\le 1.5\times 10^{-13}$, which is possible using a rubidium standard. For
instance, the
Rubidium standard developed for the Galileo satellites
has a time deviation of $10$ ps after $100$ seconds
and $1$ ns after $100000$ seconds, which corresponds to a drift rate
about $10^{-14}$. 
To keep the angular resolution below $1$ degree during 
the one year period requires
a separation distance $D\ge 4.8$ km for $5$ MHz observation frequency and
$k=1.5\times 10^{-13}$.

As can be understood by this analysis, the observations should begin with
the high frequencies, when the spacecraft are close together and the 
accumulated time error is small. As the spacecraft drift apart one should
observe at lower frequencies and take advantage of the increased 
separation distance to keep the resolution within specification.

In reality the FIRST Explorer will consist of more than two spacecraft.
The above analysis is therefore not all conclusive. However, it shows
that the concept works and the reasonable numbers so obtained have been
used in the first iteration of the design study. The design
parameters for FIRST Explorer
are summarised in Table \ref{tab:design-parameters}.

\begin{table}[b]
\caption{Design parameters for FIRST Explorer}
\label{tab:design-parameters}
\begin{tabular*}{0.9\columnwidth}{@{\extracolsep{\fill}}|l|l|l|}
\hline
{\bf{Parameter}} & {\bf{Symbol}} & {\bf{Range}}\\
\hline
\hline
Frequency & $f$ & $0.5$ -- $50$ MHz\\
\hline
Distance & D & $0.5$ -- $30$ km\\
\hline
Resolution & $\Delta\alpha$ & $\le 1$ degree \\
\hline
Time error & $\Delta t$ & $\le 1$ ns in $24$ hours\\
\hline
Range accuracy & $\Delta D$ & $\le 150$ mm at all times \\
\hline
Sensitivity & $S_\mathrm{min}$ & $\le 10$ Jy in $1$ hour \\
\hline
Bandwidth & $\Delta f$ & $100$ kHz \\
\hline
Integration time & $\Delta T$ & $1$ s -- $1$ year\\
\hline
\end{tabular*}
\end{table}
To determine the required number of spacecraft we must specify a sensitivity
requirement. Sensitivity is measured in Jansky [Jy]. It is a non-SI
unit of electromagnetic energy flux density. In SI units $1$ Jy =
$10^{-26}$Wm$^{-2}$Hz$^{-1}$.
From this definition one realises that the amount of collected energy
depends on the collecting area of the antenna, frequency bandwidth, and
observation time. The sensitivity
could in principle be lowered indefinitely by time integration but as
shown above that
is not the case due to the accumulated time error.

Intermittent bursts of low-frequency radiation are
common in the solar system and will disturb the low-frequency survey
if those emissions are not handled properly. The FIRST Explorer must be
able to blink to avoid glare and blinking requires action in
a relatively short time period.
Specifically, using a $\Delta f=100$ kHz frequency bandwidth,
the telescope shall be able to detect all non-thermal planetary radio
emission after $\Delta T=1$ hour integration time and strong
emissions from the Sun, Jupiter, Earth, and Saturn within
$1$ second. Solar type III burst and 
AKR from Earth are the strongest, in
the order of $10^8$~Jy. The Jovian radio emissions
are about $10^6$~Jy, while those from Saturn
circa $10^4$ Jy. Then follows 
Uranus and Neptune, about $100$ Jy and $10$ Jy, respectively. 

\begin{figure}[t]
\includegraphics[width=\columnwidth]{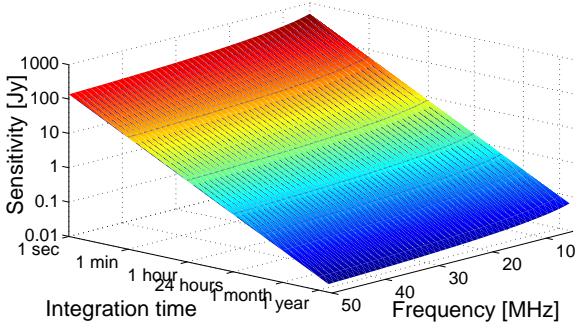}
\caption{
Sensitivity of the FIRST Explorer with six spacecraft as a function of
integration time and observation frequency.}
\label{fig:sensitivity}
\end{figure}

For any receiver system, the minimum energy flux density that can be
detected is given by
\begin{equation}
S_\mathrm{min}=\frac{S_\mathrm{sys}}{\sqrt{\Delta f \Delta T}}
\label{eq:sensivity}
\end{equation}
where $S_\mathrm{sys}=S_\mathrm{rec}+S_\mathrm{sky}$.
The sky 
is very bright at low frequencies. It has a peak about $3.6$ MHz where
the corresponding sky-noise brightness temperature, $T_\mathrm{sky}$,
reaches $T_\mathrm{max}=10^6$ K.
We can therefore 
disregard any receiver noise and set 
$S_\mathrm{sys}=S_\mathrm{sky}$ in Eq.~(\ref{eq:sensivity}).
For a short dipole antenna
$S_\mathrm{sky}=2k_\mathrm{B} T_\mathrm{sky}/A_\mathrm{dipole}$ so that
\begin{equation}
S_\mathrm{min}=
\frac{2k_\mathrm{B} T_\mathrm{sky}}{\sqrt{\Delta f \Delta T}A_\mathrm{dipole}}\,.
\label{eq:sensivity2}
\end{equation}
The effective area of a (Hertzian) dipole is 
$A_\mathrm{dipole}=3\lambda^2/8\pi$ but
between $1$ MHz and $100$ MHz, $T_\mathrm{sky}$ is empirically found to be proportional to $\lambda^{2.55}$ [Cane 1978], \ie, 
$T_\mathrm{sky}=T_\mathrm{max}(\lambda/\lambda_\mathrm{max})^{2.55}$, with
$\lambda_\mathrm{max}=83$ m. Hence
$S_\mathrm{min}$ will vary as $\lambda^{0.55}$ in this frequency range.
The FIRST Explorer formation will consist of $N$ spacecraft where each
spacecraft has a vector antenna. The dipole area in Eq.~(\ref{eq:sensivity2})
must therefore
be replaced by the corresponding expression for an $N$ element array
of vector antennas, which we take as 
$A_\mathrm{FIRST}=3 N A_\mathrm{dipole}$. 
The factor $3$ comes
from the assumption that the radiation is isotropic.
Effectively there are three $N$ element dipole arrays.

Substituting in Eq.~(\ref{eq:sensivity2}) we arrive
at the following formula
\begin{equation}
S_\mathrm{min}=\frac{16 k_\mathrm{B} T_\mathrm{max}}
{9N\sqrt{\Delta f \Delta T}}
\frac{\lambda^{0.55}}{\lambda_\mathrm{max}^{2.55}}
\label{eq:sensivity3}
\end{equation}

The analysis has shown that six spacecraft are sufficient to 
fulfil the stipulated sensitivity
requirement. 
In Fig. \ref{fig:sensitivity}, we have plotted
the sensitivity for frequencies, $f=c/\lambda$, between $5$ MHz and $50$ MHz
and integration times, $\Delta T$, between $1$ second and $1$ year,
with $N=6$ spacecraft and $\Delta f=100$ kHz bandwidth.
As can be seen, the sensitivity has a quite flat frequency responce.
For $1$ second
observations the sensitivity is in the $140-490$ Jy range  and
for $1$ hour observation in the $2-8$ Jy range.
for very long observations, $1$ -- $2$ years, the sensitivity drops
to below $90$ mJy ($f=5$ MHz, $\Delta T=1$ year).

\section{Discussion and Conclusion}
The overall conclusions from the FIRST Explorer mission concept study are
that
\begin{itemize}
\item 
FIRST Explorer can operate as a distributed aperture low-frequency radio
telescope with $1$ mother and $6$ daughter spacecraft with average
separation distances ranging from $5$ -- $30$ km in the limit, and the science
objectives will actively benefit from the daughters being allowed to
drift slowly apart in position and orientation, in an L2 or similar
stable orbit.
\item 
A low cost radio frequency range transponder can use adapted COTS mobile
phone technologies, and provide inter-spacecraft ranging to circa
$\pm 150$ mm uncertainty. In addition, the transponders can double up as the
inter-spacecraft data communications system.
\item 
Using a metrology model developed for this study, the signals from $6$
range transponders on each of the $7$ spacecraft can dynamically
monitor the formation's mutual separations and orientations in
6 DOF; enhancing the basic transponder
sensitivity by at least $15$ times to $\sim\pm 10$ mm. This provides the
ability to optimally process the science signals from the distributed
aperture array, operating at frequencies between $500$ kHz and $50$ MHz,
and achieve the required sensitivity and an angular resolution better
than $1$ degree.
\item 
All the main elements of the mission architecture, mass, power, physical
structure, data management and communications, launch and formation at
L2 have been examined and specified in sufficient detail to show that a
mission concept is basically feasible without the need to equip each
daughter spacecraft with active 6-DOF attitude and position control.
Useful science gathering operations of at least 1 to 2 years appears
feasible.
\end{itemize}

Potential key benefits of the mission and the study are that

\begin{itemize}
\item 
FIRST Explorer can provide a unique insight into the
low-frequency Universe at frequencies not available to Earth based
radio telescopes. It could also give more detailed maps of the low-frequency
Sun
and perform long-term
observations of non-thermal planetary radio emissions, providing dynamic radio
spectra of all the radio planets and may even be able to prove
the dark ages hypothesis. The mission concept therefore stands 
on its own against
competitor space science missions. 
\item 
FIRST Explorer can be a pathfinder for the concepts needed to plan 
future ``grand
science'' missions, with larger formations that could image the early
universe and directly detect Earth-like extra-solar planets,
providing valuable complementary science benefits above other more
expensive mission concepts.
\item 
The novel aspect of the FIRST Explorer is the unique combination of
existing technologies to offer the prospect for a cost-effective
multi-spacecraft science mission together with advanced metrology
modelling concepts, and the approach could be translated to enable
other science missions that use multiple spacecraft.
\item 
The study has proposed a novel idea for gross control of spacecraft
drift caused by small gravity gradients and
differential solar radiation pressure, using ``smart stability design''
and simple passive control of the effective area of the spacecraft,
with miniature solar sails.
\item 
Finally a ``metrology model'' has been developed that can predict the
positional and orientation uncertainty (in 6 DOF) of an arbitrary
constellation of objects equipped only with ranging sensors, that is
dynamically drifting and evolving in time. This tool is likely to have
applications in planning other multi-spacecraft missions.
\end{itemize}

\bigskip

\section{Acknowledgements}

This paper is developed with the support of the authors organizations
and companies, and is based on work funded by ESA under contract
Contract No. 19030/05/NL/PA. Additional thanks go to Prof. G.
Woan and Dr. T.~D.~Carozzi, Glasgow Univ., Prof. C.~McInnes and 
Dr. M.~Macdonald, 
Strathclyde Univ., Prof. P.~Wilkinson and colleagues,
Manchester Univ., and Prof. S.~Anantakrishnan, Pune Univ., 
India, for their helpful advice and
comments on the ESA study findings. We also acknowledge Dr.~James~C.~Novaco
and the 
American Astronomical
Society (AAS) for granting us permission to reproduce the 
RAE-2 all-sky image
in Fig.~\ref{fig:rae2}.

\bibliographystyle{apsrev}
\bibliography{ceas}
\end{document}